\documentclass[slac_one]{revtex4}
\usepackage{graphicx}
\usepackage{fancyhdr}
\begin{document}


\title{{\small{2005 International Linear Collider Workshop - Stanford,
U.S.A.}}\\ %% Please keep this conference title here
\vspace{12pt}
 Monte Carlo Study of a Luminosity Detector for the International Linear Collider\footnote{This work was partly supported by the Nathan Cummings Chair of Experimental Particle Physics.}} %% Paper title goes here


%




\author{H.~Abramowicz\footnote{also at Max Planck Institute, Munich, Germany, Alexander von Humboldt Research Award.}, R.~Ingbir, S.~Kananov and A.~Levy}
\affiliation{School of Physics and Astronomy, Raymond and Beverly Sackler Faculty of Exact Sciences, Tel Aviv University, Tel-Aviv, ISRAEL}

\begin{abstract}

This paper presents the status of Monte Carlo simulation of one of the
luminosity detectors considered for the future $e^+e^-$ International Linear
Collider (ILC).  The detector consists of a tungsten/silicon sandwich
calorimeter with pad readout. The study was performed for Bhabha
scattering events assuming a zero crossing angle for the beams.

\end{abstract}

\maketitle

\thispagestyle{fancy}

\section{Introduction}

The linear collider community has set a goal to achieve a relative
precision of $10^{-4}$ on luminosity measurement. Presently the
Forward Calorimetry Collaboration (FCAL)~\cite{fcal} is considering
two possible designs for the luminosity detector (LumiCal). Both
designs are based on a tungsten/silicon calorimeter. They differ in
the readout design, pad or strip. Here we report on studies performed
to optimize the performance of the pad readout design.

\section{Simulation scheme}

The Monte Carlo studies, presented in the current note, include
simulation of Bhabha scattering, beam-beam interactions, beam spread
as well as the full simulation of the LumiCal. For Bhabha scattering
events we used BHWIDE~\cite{bhwide}, a Monte Carlo multi-photon event
generator which provides four-momenta of the outgoing electron,
positron and photons radiated in the initial and final state. The
program CIRCE ~\cite{circe} was used to study the distortion of beam
energy spectrum due to beamstrahlung. Two different values of a
Gaussian beam spread, 0.05\% and 0.5\% of the nominal center of mass
energy ($\sqrt{s}$), were investigated in the range of beam energy
between 50 and 400 GeV . The detector simulation was performed using
the BRAHMS~\cite{brahms} package based on the standard GEANT 3.21
simulation program~\cite{geant}. The performance of LumiCal was studied in
three stages,

\begin{itemize}
\item with the basic detector design using single electrons/positrons,
\item with the basic detector design and a more realistic physics
  simulation, including simulation of Bhabha scattering events,
  beamstrahlung and beam spread,
\item varying the detector design for optimization purposes. 
\end{itemize}

\begin{figure*}[t]
\centering
\includegraphics[width=110mm,height=64mm]{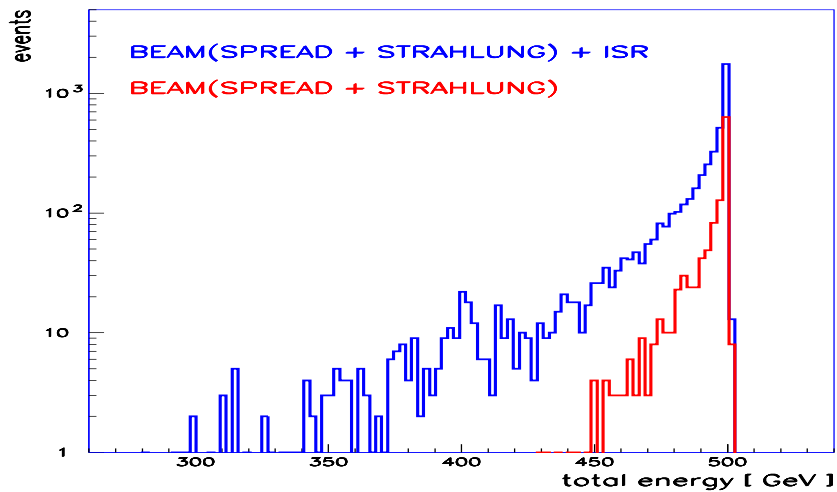}
\caption{Energy spectrum of the $e^+e^-$ using BHWIDE and CIRCE at $\sqrt{s}=500 GeV$ with a beam spread of 0.05\%$\sqrt{s}$, with and without initial state radiation, ISR, as described in the figure.} 
\label{es}
\end{figure*}

\noindent
Fig.~\ref{es} shows an example of the center of mass energy spectrum
of the $e^+e^-$ pair originating from Bhabha scattering, including
radiative effects, beamstrahlung and a beam spread of 0.05\%$\sqrt{s}$
for nominal 250 GeV beam energies. The main contribution to the tail
comes from the initial state radiation in Bhabha scattering.

\section{LumiCal Design}

The detector covers polar angles $\theta$ from 27 to 91 mrad with
respect to the beam line. Longitudinally, the detector consists of 30
layers composed each of 500 $\mu$m thick silicon sensors and a
tungsten-silicon mixture of 0.34 cm of tungsten and 0.31 cm of silicon
and electronics.  The detector, with an inner radius of 8 cm and an
outer radius of 28 cm, is subdivided radially into 15 cylinders and
azimuthally into 24 sectors.

Each layer corresponds to a depth of about one radiation length. The
cell transverse size is approximately one $Moli\acute{e}re$ radius.
The total number of cells (electronic channels) is equal to 10,800.
Two identical arms, one for the electron side and the second for the
positron side, are positioned along the $z$ axis (beam line),
symmetrically with respect to the interaction point (IP), 3.05 m away
from the IP.

\section{Event selection}

\begin{figure*}[t]
\centering
\includegraphics[width=110mm,height=64mm]{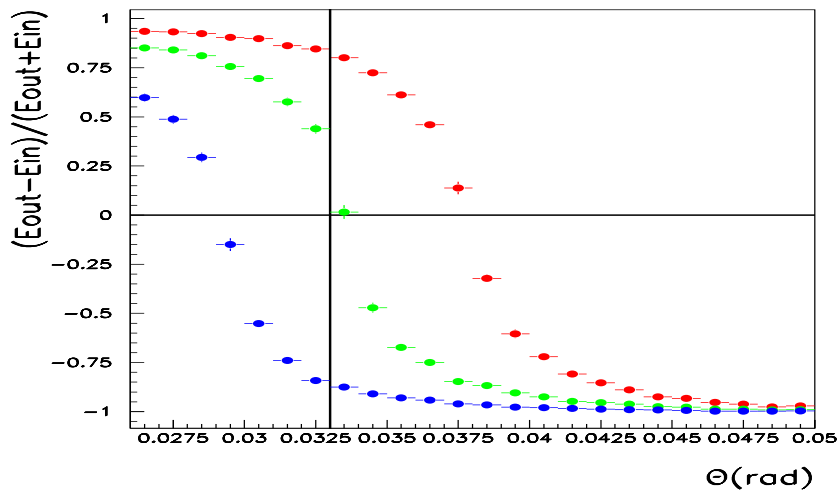}
\caption{The variable ($E_{out}$-$E_{in}$)/($E_{out}$+$E_{in}$) as 
  a function of the generated polar angle, $\theta_{gen}$, of the
  showering electrons for $E_{out}$ summed over one (blue points), two
  (green points) or three (red points) inner cylinders.}
\label{selection}
\end{figure*}

For luminosity measurement, the geometric acceptance is the most
significant event selection rule. The strong $\theta$ dependence of
the Bhabha scattering, $d\sigma/d\theta\sim 1/\theta^3$, makes the low
angle cut crucial.  We used a method, in which only a few layers
govern the events selection. In this method the energy deposited in
three layers located in the middle of the detector, close to the
shower maximum, is divided into energy deposited in the inner edge
cylinders, $E_{out}$ and outside, $E_{in}$.
The variable, {\it{p}}, defined as
\begin{equation}
p=\frac{E_{out}-E_{in}}{E_{out}+E_{in}} 
\end{equation}
is then used to estimate the shower containment.  Events with
{\it{p}}$>$0 are rejected as being out of the acceptance region, while
events with {\it{p}}$<$0 are kept.  The behavior of the variable
{\it{p}} as a function of the polar angle of the showering electron is
shown in Fig.~\ref{selection} for three different definitions of
$E_{out}$, summed over one, two, or three cylinders. A given
acceptance cut in $\theta$ can be translated into an appropriate
number of edge cylinders such that a cut on $p$ will reject the right
events without necessity for full reconstruction.
The events were preselected using {\it{p}} only calculated for the
electron detector arm. The Bhabha scattering events were further
selected by requiring that the showers reconstructed in both arms of
the detector be back to back.

\section{Position reconstruction}
 
Two approaches were used to reconstruct the position of the shower,
with no attempt to reconstruct clusters of energy. The position was
determined as the weighted average,
\begin{equation}
<x>=\frac{\sum_i x_i W_i}{\sum_i W_i} ,
\end{equation}
where $x_i$ is the location of the center of the $i$-th pad, $W_i$ is
the weight and the sum runs over all pads. The simple, energy weighted
average, $W_i=E_i$, is known to be biased, with the bias depending on
the size of the pads and the impact point of the shower on the pad.

\begin{figure*}[t]
\centering
\includegraphics[width=110mm,height=64mm]{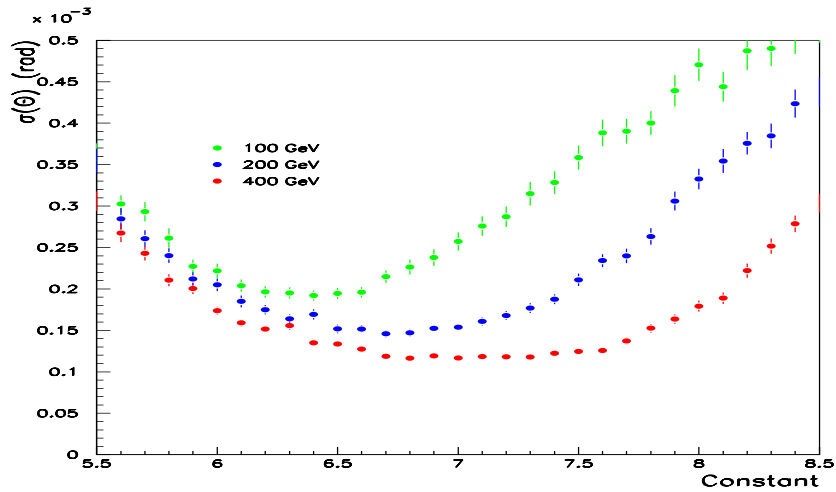}
\caption{Polar angle resolution, $\sigma (\theta)$, as a function of the cut-off value $constants$, for three beam energies, as denoted in the figure.}
\label{con}
\end{figure*}

In the second method~\cite{log_wei}, the weight is assumed to be
proportional to the logarithm of the energy deposited, and in addition
a cut-off is introduced so that effectively only significant energy
deposits contribute,
\begin{equation}
W_i=\max\{0,[const+\ln\frac{E_i}{E_{tot}}]\},
\label{logwei}
\end{equation} 
where $E_{tot}=\sum_i{E_i}$. The cut-off, $const$, has to be optimized
for best resolution. The resolution as a function of cut-off is shown
in Fig.~\ref{con} for three different incoming energies. The optimal
cut-off is found to increase with energy. In parallel, for each
cut-off value the bias in reconstructing the position was checked. The
point of best resolution turned out to correspond to the least bias.

Just by tuning the cut-off, the polar angle resolution ,
$\sigma(\theta)$, for 250 GeV energy electrons was found to improve by
factor three with $\sigma(\theta)= 0.136 \pm 0.003$~mrad.  In addition,
the bias was improved by an order of magnitude.

\section{Detector performance}

\begin{figure*}[t]
\centering
\includegraphics[width=110mm,height=64mm]{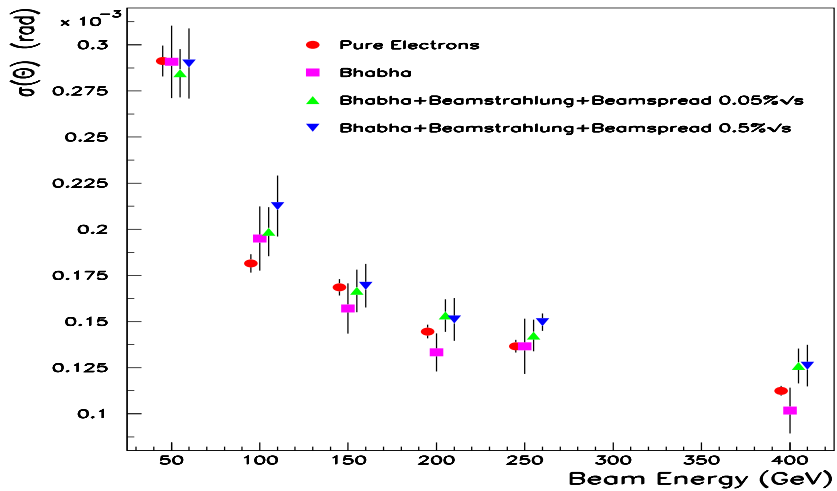}
\caption{Polar angle resolution, $\sigma (\theta)$, as a function of the 
  beam energy for different physics cases, as denoted in the figure.}
\label{thres}
\end{figure*}

The polar angle resolution obtained for single particles with the
optimal weighting, is shown as a function of the beam energy in
Fig.~\ref{thres}, for the various event configurations. Typically, the
best resolution is achieved for the single particle MC sample, while
for Bhabha scattering with a 0.5\% beam spread the resolution is 10\%
worse.
%With such resolution, to achieve the required level of precision on
%luminosity, it would have to be known to within 1\%. Obviously, this
%will be very difficult to achieve and further improvements are
%necessary.

A small residual bias in the polar angle reconstruction, $\Delta
\theta$, was observed. For Bhabha scattering with a 0.05\% beam
spread, the relative value $\Delta \theta/\theta_{min} =
(5.7\pm1.3)\cdot 10^{-4}$ is of the same order of magnitude as the
required luminosity precision. More statistics is needed to establish,
whether this is a genuine effect.

\begin{figure*}[t]
\centering
\includegraphics[width=110mm,height=64mm]{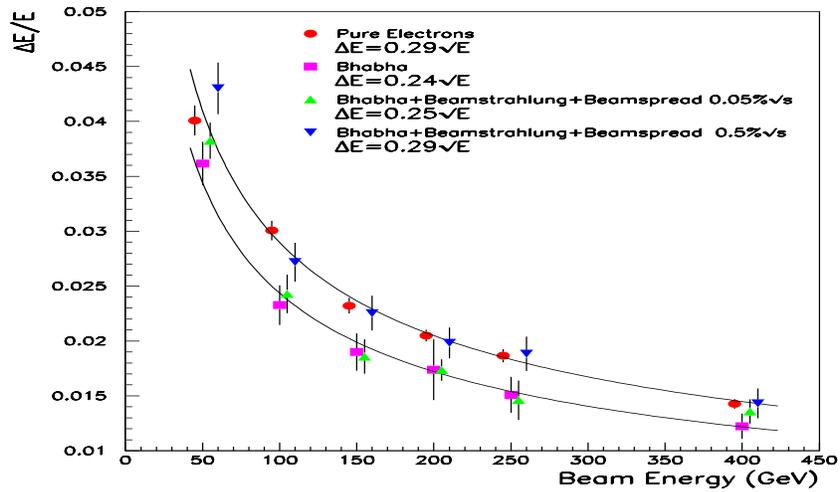}
\caption{Energy resolution, $\Delta E$, as a function of energy for
 different physics cases, as denoted in the figure.}
\label{eres}
\end{figure*}

The energy resolution, $\Delta E$, as a function of energy is shown in
Fig.~\ref{eres}, again for the various event samples. The resolution
follows the expected behavior of $\Delta E/E = a/\sqrt{E}$, with $a$
varying between 0.24$\pm$0.02 and 0.29$\pm$0.03. The best resolution
is achieved for the sample of Bhabha scattering events with a
small beam spread. This is probably due to the requirement that the
two showers be back to back, which prevents residual energy leakage.

\section{Design Optimization}

Once the performance of the basic detector design was established , an
attempt was made to optimize the depth and the granularity of the
calorimeter. For that purpose the calorimeter was assumed to have 50
active layers. A sample of 1000 events was generated for this study.

\begin{figure*}[t]
\centering
\includegraphics[width=110mm,height=64mm]{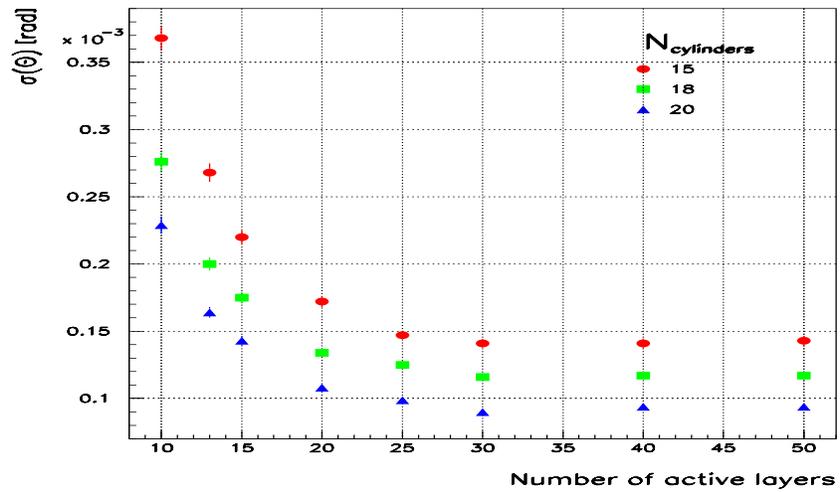}
\caption{
  The polar angle resolution, $\sigma(\theta)$, as a function of the
  detector length expressed in terms of active layers, for varying
  number of cylinders (radial granularity).}
\label{dns}
\end{figure*}

The angular resolution was studied as a function of the depth and
granularity, which was improved by increasing the number of cylinders
for the same geometry. The results are shown in
Fig.~\ref{dns}. For a given depth of the calorimeter, the
resolution is improving with increasing number of readout pads.  No
improvement is observed beyond a depth of 30 layers. For 30 active
layers, increasing the number of cylinders from the 15 of the basic
design to 20, leads to an angular resolution better than $10^{-4} $ rad.

The number of sectors was also increased. This improves the resolution
of the azimuthal angle, but has no effect on the polar resolution.
Increasing the density of the calorimeter would lead to a more compact
detector, with less leakage. However the number of pads would have to
be increased to match the angular performance of the basic design .

An attempt was made to achieve a better angular resolution by
improving the granularity locally, in the layers containing the
maximum of the electromagnetic shower. If this is done under the
constraint of a fixed total number of readout channels, with 20
cylinders in the inner 15 layers and 10 cylinders in the outer layers,
the resolution may be improved by 15\%.

\section{High statistics MC}

To achieve sensitivity to systematic effects comparable to the
required relative precision on luminosity of $10^{-4}$, large
statistics MC samples are necessary. This cannot be achieved in a
conventional manner, by processing events through a full GEANT
simulation. Instead, a fast MC was developed, with smearing effects
implemented through parameterization of the performance established on
smaller samples. This MC allows detailed studies of various
systematic effects, either related to geometry or possible mismatch
between MC simulation and detector performance in reality.

\begin{figure*}[t]
\centering
\includegraphics[width=110mm,height=64mm]{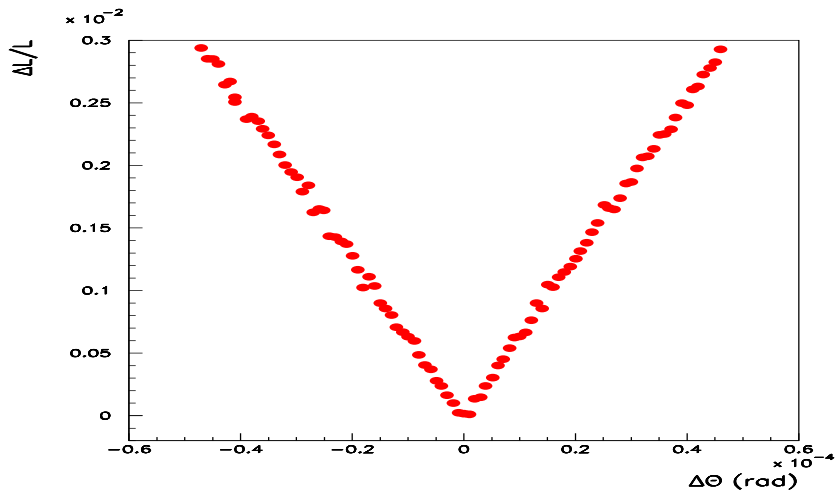}
\caption{The relative error on the luminosity, $\Delta L/L$, as a function of the assumed bias in the polar angle reconstruction, $\Delta \theta$. }
\label{delta-l}
\end{figure*}

The influence of a bias in $\theta$ reconstruction on the luminosity
error, obtained with fast MC simulation, is shown in
Fig.~\ref{delta-l}. Sensitivity to shifts of the order of $10^{-4}$
are visible. The simulation reproduces well the expected analytical
result, $\Delta L/L = 2\Delta \theta/\theta_{min}$.

The fast MC was also used to assess whether the presently achieved
angular resolution would be sufficient to control the $\Delta L/L$
with the required precision. Assuming that the Bhabha scattering
events are selected by requiring that the two electromagnetic showers
be back to back within three times the expected resolution, and
assuming a precision of 10\% on the resolution itself, would lead to
an error on the luminosity, $\Delta L/L \simeq 5\cdot
10^{-4}$. However, if the back to back requirement is relaxed to five
times the expected resolution, the projected luminosity error, for the
same uncertainty on the resolution, is negligible. Therefore, further
studies are needed to understand the required quality for selecting
Bhabha candidate events.

\section{Summary}

The luminosity detector at the future linear collider is expected to
provide measurements with a precision better than $10^{-4}$. In this
study, of a tungsten-silicon calorimeter with pad readout, we have
concentrated mainly on optimizing the reconstruction algorithms.  We
have demonstrated that an improvement of a factor of three is possible
without changing the granularity. An attempt was also made to optimize
the angular resolution by changing the granularity, either by
increasing the total number of readout channels or by improving the
granularity at the expected shower maximum location while keeping the
total number of channels unchanged. A fast MC was developed to study
systematics effects with a sensitivity compatible with the required
precision.

\end{document}